\documentclass{article}

\usepackage{arxiv}

\usepackage[utf8]{inputenc} % allow utf-8 input
\usepackage[T1]{fontenc}    % use 8-bit T1 fonts
\usepackage{hyperref}       % hyperlinks
\usepackage{url}            % simple URL typesetting
\usepackage{booktabs}       % professional-quality tables
\usepackage{amsfonts}       % blackboard math symbols
\usepackage{nicefrac}       % compact symbols for 1/2, etc.
\usepackage{microtype}      % microtypography
\usepackage{lipsum}		% Can be removed after putting your text content
\usepackage{graphicx}
\PassOptionsToPackage{numbers,sort&compress}{natbib}

\usepackage{natbib} % もしまだ呼んでいなければ
\usepackage{doi}
\usepackage{amsmath}
\usepackage{CJKutf8}
\usepackage[subrefformat=parens]{subcaption}
\usepackage{here}
\usepackage{booktabs}
\usepackage{graphicx}

\usepackage[right]{lineno}
\usepackage{bm} 
\modulolinenumbers[5]
\title{A Simulation Study on the Cosmic Ray Energy Spectra of Elemental Mass Groups using the Tibet Air Shower and Muon Detector Arrays through the Bayesian Unfolding Method}

\date{}	% Here you can change the date presented in the paper title
\author{
\begin{tabular}{c}
G. Imaizumi, M. Anzorena, K. Fujita, T. Kawashima, K. Kawata, 
A. Mizuno,\\ M. Ohnishi, R. Garcia, T. Sako, F. Sugimoto, 
M. Takita, Y. Yokoe \end{tabular} \\
Institute for Cosmic Ray Research, University of Tokyo, Kashiwa, Japan \\
\texttt{ginga@icrr.u-tokyo.ac.jp}
\AND
Y. Katayose \\
Faculty of Engineering, Yokohama National University, Yokohama, Japan
\AND
S. Kato \\
Institut d’Astrophysique de Paris, Paris, France
}
\renewcommand{\shorttitle}{A Simulation Study on the Cosmic Ray Energy Spectra of Elemental Mass Groups using the Tibet Air Shower and Muon Detector Arrays through the Bayesian Unfolding Method}

\begin{document}
\maketitle

\begin{abstract}
\noindent
We study the analysis method to determine the cosmic-ray energy spectra of different mass groups assuming the use of the Tibet AS$\gamma$ experiment, which consists of the high-density Tibet air-shower array (Tibet-AS) and the underground muon detector (MD) array. These arrays measure the sampling air shower size $\Sigma\rho$ and the total muon number $\Sigma N_{\mu}$ of each air shower event. These parameters are known to contain information on the energy and mass of the primary particle. To reconstruct the energy spectra of individual cosmic-ray mass groups, we apply a multidimensional unfolding method based on Bayes' theorem to the two-dimensional distribution of $\Sigma\rho$ and $\Sigma N_{\mu}$ produced by Monte Carlo simulation. Simulated datasets with combinations of the EPOS-LHC, SIBYLL-2.3c, and QGSJET~II-04 high-energy hadronic interaction models and a helium-dominant composition model are analyzed while using a response matrix produced by EPOS-LHC.
The unfolded spectra of the EPOS + helium-dominant composition model dataset show a deviation from the input flux within $\pm$10\% except for a few bins, meaning that the uncertainty of the technique itself and the composition model dependence is at that level.
It is also shown that the deviation in the all-particle spectrum is within $\pm$10\% even when using different hadronic interaction models in the dataset and the response matrix.
On the other hand, the unfolded spectra of individual mass groups have a clear dependence on the hadronic interaction model.
The model dependence of the proton and helium spectra amounts to $\pm25$\% below 10$^{6.5}$ GeV.
The dependence in the carbon group is at a $\pm25\text{\%}$ level below 10$^{6}$ GeV, and for the iron spectrum, it amounts to $+55\%$ and $-30\%$ in the energy range of 10$^{5.1}$ GeV to 10$^{6.7}$ GeV.
We are now able to measure individual energy spectra in the knee region—including those of heavy nuclei, which remain poorly understood. Thanks to the recent extension of direct measurements beyond 10$^{4}$ GeV, a comparison between our analysis and the direct measurements will provide a good test of the hadronic interaction models.

\end{abstract}

\section{Introduction}
\noindent
The steepening feature of the energy spectrum of cosmic rays around $10^{15.5},\textrm{eV}$ is known as the knee \cite{M. Nagano.}.
It is usually explained as the acceleration limit of cosmic-ray protons in our galaxy, but the energy spectrum structures of each nuclear group in this energy range are still an open question.
Although a simple scenario of the galactic cosmic rays predicts a proton-dominant composition below the knee energy, to understand these spectral structures—hence the origin and propagation of the galactic cosmic rays—measurements of the energy spectra of individual mass groups below the knee energy become important.
The recent progress in direct measurements by CALET, DAMPE and CREAM reporting a hardening of the proton spectrum at a few hundred GeV and a softening starting around 10 $\sim$ 20 TeV \cite{CALET CollaborationP, DAMPE CollaborationP,CREAMPandHe} supports a complex structure around 10$^{13}$ eV.
%Moreover, they showed the helium spectrum experiences a hardening at $\sim$ TeV energies and then shows a softening above a few tens of TeV\cite{CALET CollaborationHe,DAMPE CollaborationHe}.
Above 10$^{13}$\,eV, the nuclear mass composition or energy spectra of each mass group are measured indirectly by air-shower experiments.

Indirect measurements conducted in the HAWC experiment found that the all-particle energy spectrum has a softening at 40\,TeV \cite{R. Alfaro allspe}, which is consistent with the softening of the light elements introduced above.
The GRAPES-3 experiment suggested that the proton flux has another hardening at 1.66$\times$10$^{14}$ eV \cite{GRAPES-3 Collaboration}.
Also, recent LHAASO measurements of $\langle \mathrm{ln} A \rangle$ suggest a light composition in the 10$^{14.5-16}$ eV range \cite{LHAASO Collaboration,{LHAASO lnA}}.
There are results for all particles and light particles. But individual results including heavy particles are not yet available around the knee.
The extension of the energy coverage by direct measurements beyond the 10$^{13}$\,eV region may allow us to compare the direct and indirect measurements, which will constrain the uncertainty of the hadronic interaction models.

In this study, we investigate the performance of energy spectrum determination in some mass groups by Monte Carlo simulation, assuming the use of the Tibet AS and muon detector (MD) arrays.
Using the surface detector array (AS) and MDs, we can measure the number of charged particles ($\Sigma \rho$) and the number of muons ($\Sigma N_{\mu}$), respectively, event-by-event at the observation height.
These observables are sensitive to the energy and mass of the primary particle, as demonstrated by Matthews \cite{J. Matthews}.
An unfolding technique of the two observables ($\Sigma \rho$ and $\Sigma N_{\mu}$) to the primary information (energy and mass number A) was developed by the KASCADE and KASCADE-Grande experiments \cite{KASCADE-GrandeD,KASCADE} and applied in the analysis above the knee \cite{KASCADE-Grande}.
In this study, we apply this technique to the Tibet experiment simulation data.
Because the Tibet experiment is located at 4,300 m above sea level, in contrast to the KASCADE experiment operated at 110\,m a.s.l., our study has the best sensitivity at lower energies, complementary to the KASCADE measurement.
In addition, as the MDs have a total 3,400 m$^{2}$ detection area and are installed below a 2.4 m soil overburden, we detect a sufficient number of muons with high purity even at lower energies.

This paper is organized as follows.
In Sec.\ref{sec:detector} the experimental setup of the Tibet AS and MD arrays is explained.
Sec.\ref{sec:monte-carlo} describes the setup of the air shower and detector simulations, followed by an explanation of the event reconstruction and data selection in Sec.\ref{sec:shower-analysis}.
After defining the formulae of the unfolding procedure in Sec.\ref{sec:unfolding} the results of the unfolded spectra are discussed in Sec.\ref{sec:results}
Sec.\ref{sec-conclusion} concludes the paper.
%%%%%%%%%%%%%%%%%%%%%%%%%%%%%%%%%
%%% Section : Tibet AS and MD %%%
%%%%%%%%%%%%%%%%%%%%%%%%%%%%%%%%%
\section{Tibet Air shower and MD arrays} 
\label{sec:detector}

\begin{figure}[htbp]
	\begin{center}
		\includegraphics[width=0.7\linewidth]{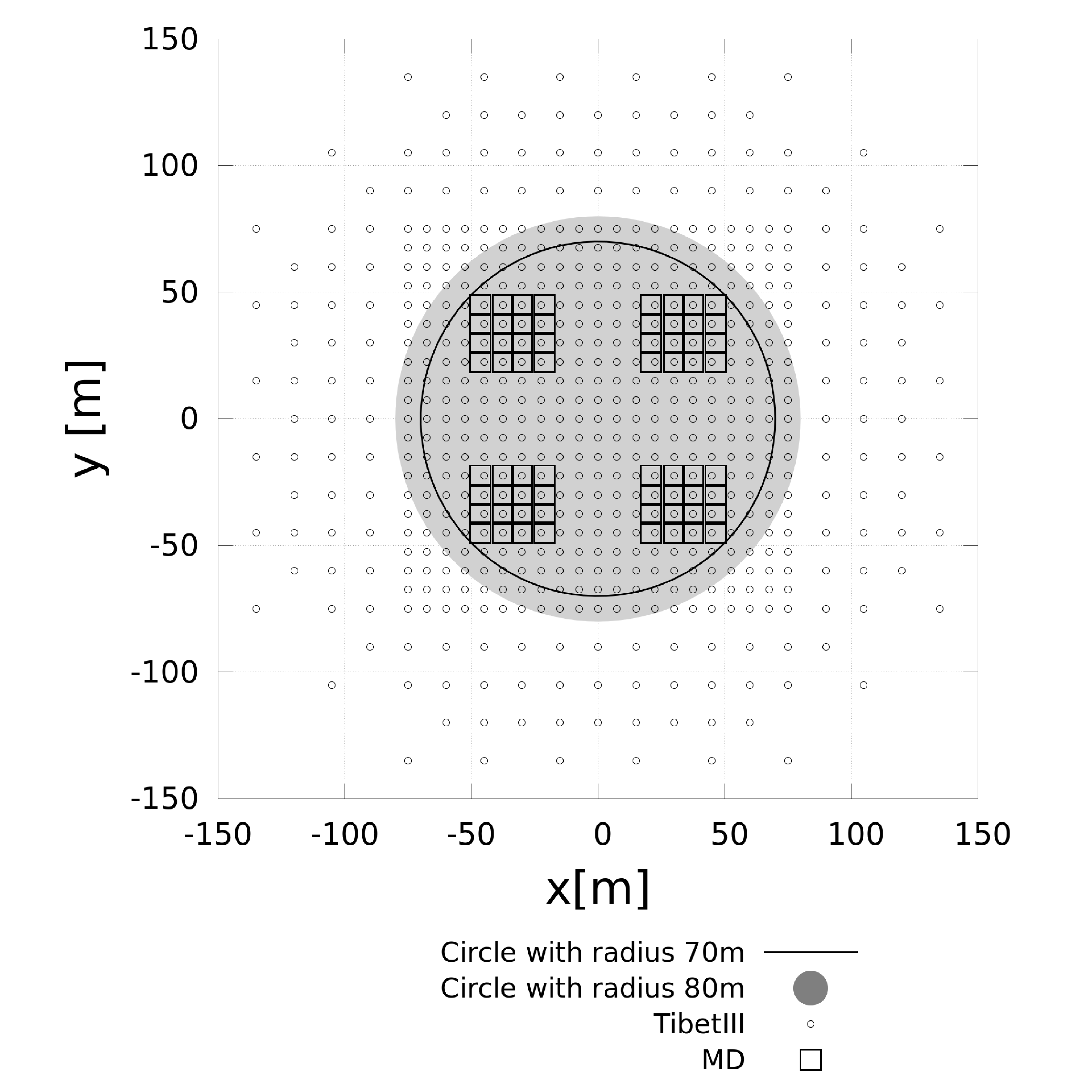}
	\end{center}
\caption{Top view of the Tibet air-shower (AS) array and the underground muon detector (MD) array. Small open circles represent 0.5\,m$^2$ scintillation counters that make up the Tibet-AS. Open squares denote MD pool cells, which are installed beneath 2.4\,m of soil and a 0.2\,m-thick concrete layer. The area within an 80\,m radius from the center of the array is indicated by gray hatching\protect\footnotemark[1]. A solid circle indicates the geometrical detection area with a 70\,m radius\protect\footnotemark[2].}

 		\label{fig:Tibet-III+MD}
\end{figure}

\begin{figure}[htbp]
	\begin{center}
		\includegraphics[width=0.7\linewidth]{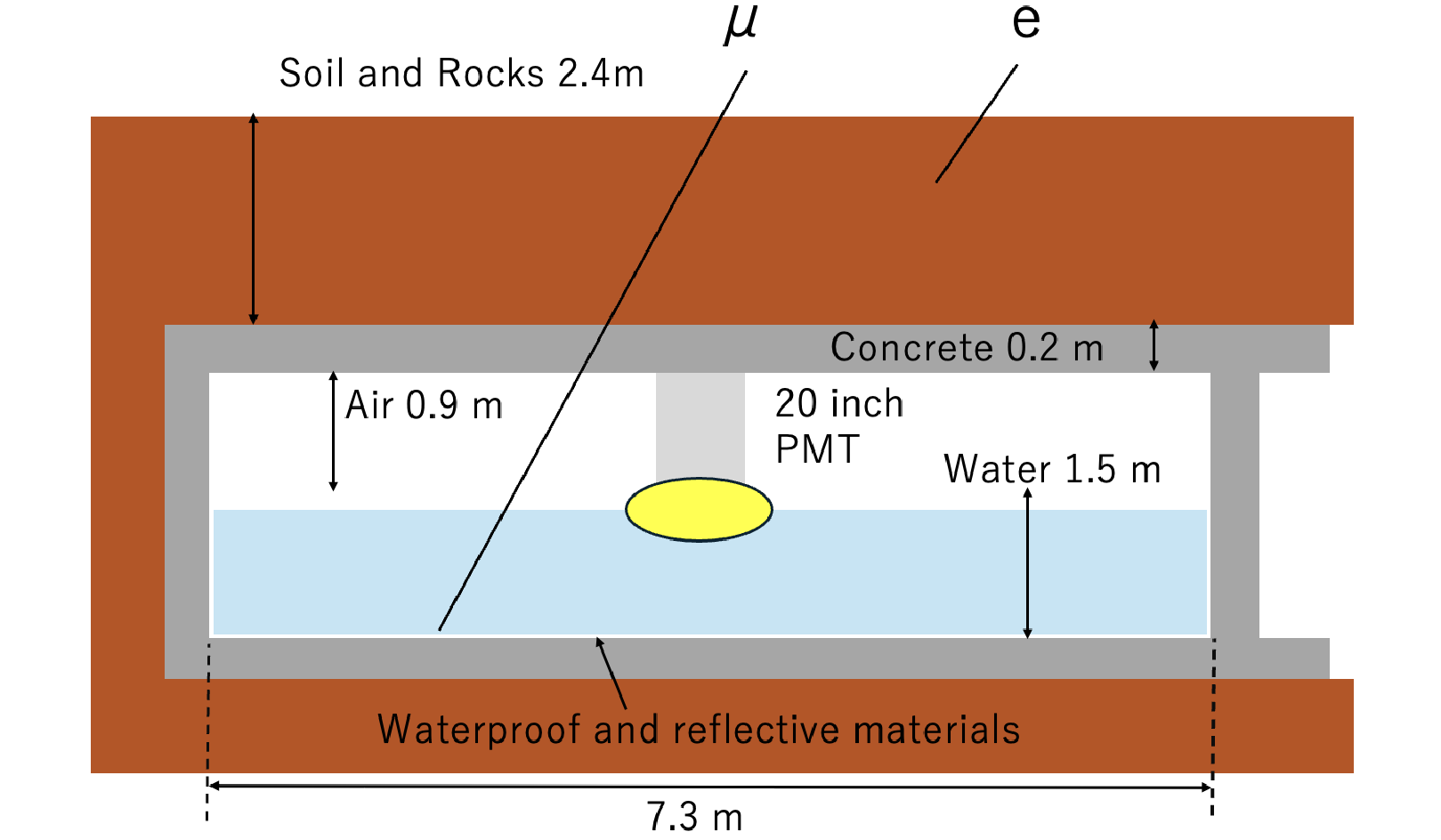}
	\end{center}
\caption{Schematic view of a MDcell. Each pool measures 7.3\,m $\times$ 7.3\,m and is filled with 1.5\,m of water. A photomultiplier tube (PMT) is mounted on the ceiling of each cell.}

 		\label{fig:MD}
\end{figure}

\noindent
The Tibet AS and MD arrays have been observing very-high-energy cosmic rays at Yangbajing ($90.522^{\circ}$ east, $30.102^{\circ}$ north, 4,300 m a.s.l., 606 g/cm$^2$ atmospheric depth) in Tibet, China \cite{M. Amenomori allspe}.
The current AS array, covering an area of 65,700 m$^2$, consists of 597 plastic scintillation detectors indicated by small open circles in Fig.\ref{fig:Tibet-III+MD}, each with a 0.5 m$^2$ detection area.
With a 5 mm-thick lead plate placed above each scintillation detector, this array detects the electromagnetic component—including electrons, positrons, and gamma rays—in the air showers and measures the arrival timing and the density of air-shower particles at each detector.
The arrival direction of the primary cosmic ray is reconstructed from the relative timings of the hit detectors.
The MD array consists of 64 water Cherenkov-type cells placed 2.4 m beneath Tibet-AS.
Each MD cell is a concrete water tank with an area of 7.3 m $\times$ 7.3 m filled with clean water to a depth of 1.5 m and equipped with a 20-inch-diameter photomultiplier tube (PMT) mounted downward on the ceiling, as shown in Fig.\ref{fig:MD}.
The inner surface of the MD cell is waterproofed and covered with a white Tyvek sheet to reflect Cherenkov light, which is emitted by air-shower muons.

%%% Fig.-1 %%%%%%%%%%%%%%%%%%%%%%%%%%%%%%%%%%%%%%%%%%%%%%%%%

\footnotetext[1]{Hit detectors area. As explained in Sec. ~\ref{sec:event-selection}}
\footnotetext[2]{The air shower core area. As explained in Sec. ~\ref{sec:event-selection}}
%%% Fig.-2 %%%%%%%%%%%%%%%%%%%%%%%%%%%%%%%%%%%%%%%%%%%%%%%%%

%%%%%%%%%%%%%%%%%%%%%%%%%%%
% Section : Monte Carlo %%%
%%%%%%%%%%%%%%%%%%%%%%%%%%%
\section{Monte Carlo Simulations for Air Shower and Detector Response}
\label{sec:monte-carlo}
\noindent
To develop a reconstruction method for the primary cosmic-ray energy spectrum of each mass group from the number of charged particles and the number of muons obtained by the Tibet AS and MD arrays,
we performed Monte Carlo simulations of the air-shower development and detector response.

The air-shower generation in the atmosphere was performed using the CORSIKA (version 7.6400) package \cite{D. Heck},
while the responses of the scintillation detectors and the muon detectors were simulated using the GEANT,4.10.02 toolkit \cite{GEANT4 Collaboration}.
In the CORSIKA simulation, we used the hadronic interaction models EPOS-LHC \cite{Pierog}, SIBYLL-2.3c \cite{Riehn Felix} and QGSJET II-04 \cite{Ostapchenko S} above 80 GeV to estimate the systematic uncertainties due to the choice of interaction model.
For the hadronic interaction model below 80\,GeV, we used the FLUKA model \cite{Fasso,Fluka}.
% including the Landau–Pomeranchuk–Migdal effect and a special treatment of the tip of the bremsstrahlung spectrum

\begin{figure}[htbp]
\begin{minipage}[b]{0.55\linewidth}
\begin{center}
 \includegraphics[width=0.9\linewidth]{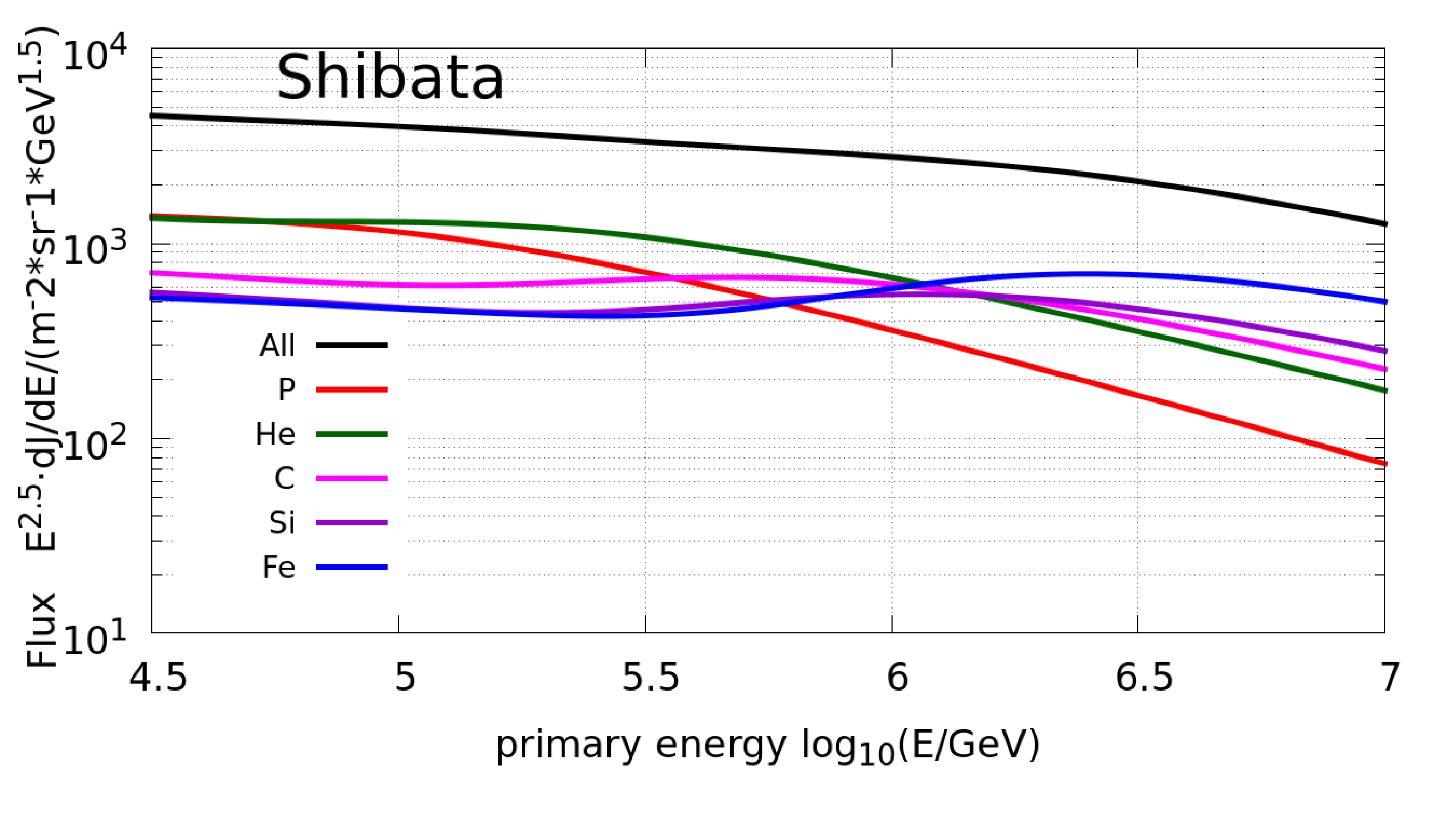}
\end{center}
 \subcaption{}
\end{minipage}
\begin{minipage}[b]{0.55\linewidth}
\begin{center}
\includegraphics[width=0.9\linewidth]{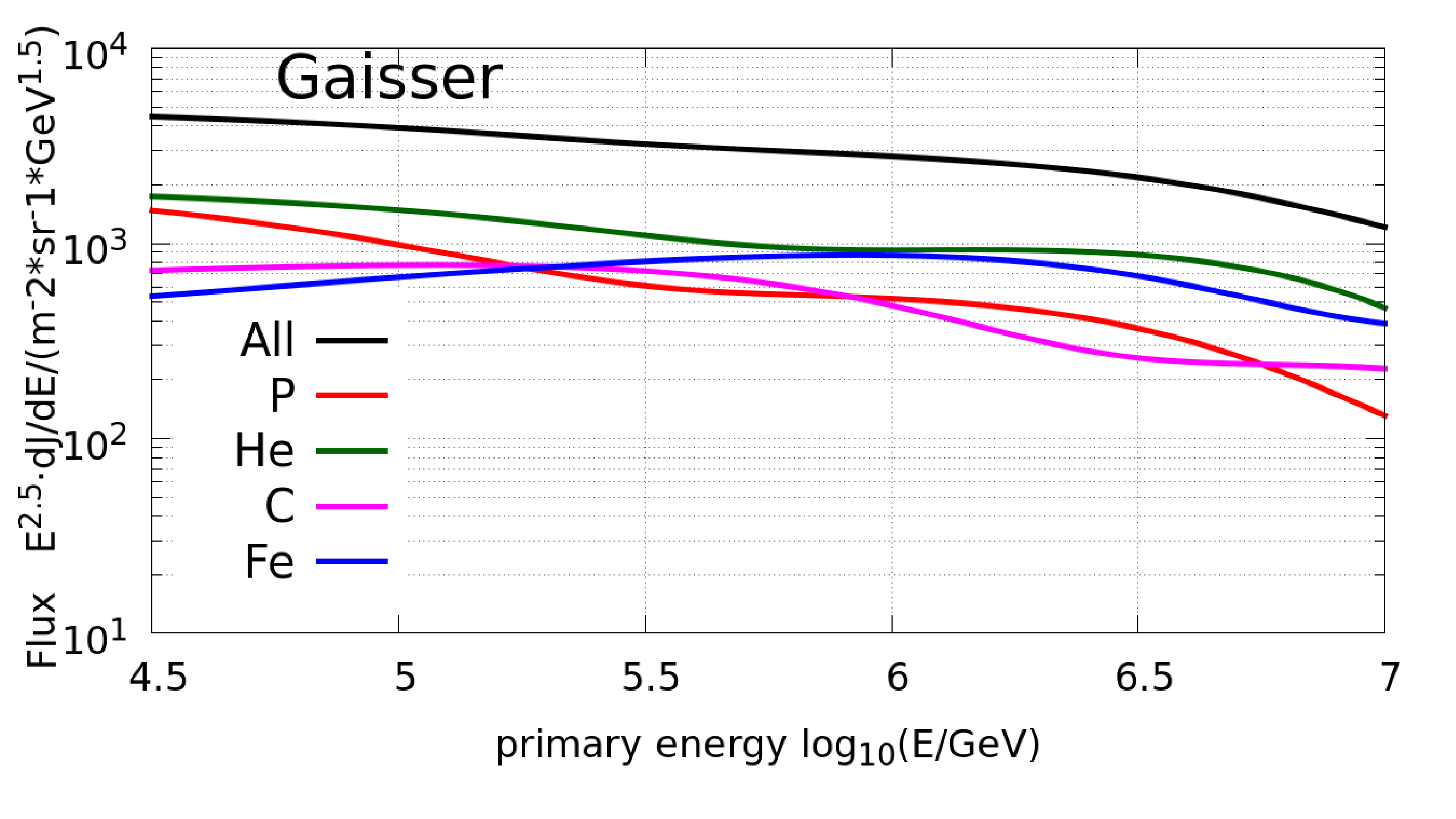}
\end{center}
 \subcaption{}
\end{minipage}

\caption{
 Primary CR spectra of each mass in (a) the Shibata model and
(b) the Gaisser model.
}
\label{Fig:energyspe}
\end{figure}

As the primary cosmic-ray spectrum, we used two mass composition models.
One is the Shibata model \cite{M. Shibata}, as shown in Fig.\ref{Fig:energyspe} (a), which consists of protons (P), helium (He), carbon group (C\protect\footnotemark[3]),
silicon group (Si\protect\footnotemark[4]), and iron (Fe\protect\footnotemark[5]).
In this work, we used only P, He, C, and Fe.
In this model, the heavier mass components gradually dominate above 100 TeV.
The other is the Gaisser model \cite{T. K. Gaisser} as shown in Fig.\ref{Fig:energyspe} (b), which consists of P, He, C, and Fe.
In this model, the He component gradually dominates above the tens of TeV region.

All air-shower events were generated at zenith angles of $0^\circ \le \theta \le 45^\circ$ and within a radius of 300 m from the array center.
To ensure sufficient statistics of cosmic-ray events with the steep spectral index, the air-showers were generated with two different energy thresholds:
3$\times10^{3.0} \le E \le 10^{7.0}$ GeV ($4.4 \times10^8$ events, equivalent to 0.4 days of observation) and $10^{5.0} \le E \le 10^{7.0}$ GeV ($9.0 \times10^6$ events, equivalent to 2.6 days).
The events with energies in the range 3$\times10^{3.0} \le E \le 10^{7.0}$ GeV were assigned appropriate weights to match the effective observational duration of the $10^{5.0} \le E \le 10^{7.0}$ GeV dataset.
The MC simulation settings are summarized in Table~\ref{tab:1}.

\footnotetext[3]{C represents the carbon, nitrogen, and oxygen group.}
\footnotetext[4]{The Si group represents the intermediate elemental group (between F and Mn) .}
\footnotetext[5]{Fe represents the heavy component of cosmic rays, referring to iron and elements heavier than Fe.}

%%% Fig.-6 %%%%%%%%%%%%%%%%%%%%%%%%%%%%%%%%%%%%%%%%%%%%%%%%%

\begin{table}[h]
\begin{center}
% table caption is above the general features of QCD like quark confinement, multiple interactions and jet production are included in the model,table
\caption{Simulation settings.}
\label{tab:1}       % Give a unique label
% For LaTeX tables use
\begin{tabular}{ll}
\toprule\noalign{\smallskip}
Air shower simulation & CORSIKA (version 7.6400) \\
Primary energies   & 3$\times10^{3.0} \le E \le10^{7.0}$ GeV and\\ & $10^{5.0} \le E \le10^{7.0}$ GeV  \\
 Zenith angle $\theta$   &    $ 0^\circ \le \theta \le 45^\circ$\\
Observation altitude  & 4,300\,m  \\
Minimum kinetic energy of secondaries & electrons, gamma rays, $\pi^0$: 1MeV \\
                                                       &hadrons: 1 GeV, muons : 50 MeV \\ 
Interaction models  
                      &  EPOS-LHC-v3400 and FLUKA\,2011.2x \\
                         &  SIBYLL 2.3c-01  and FLUKA\,2011.2x  \\ 
                        &  QGSJET-II-04 and FLUKA\,2011.2x \\
Cosmic-ray composition modes & Shibata model\,(Heavy dominant) \\
                              &Gaisser model (Helium dominant)\\
                              
Detector simulation & GEANT4.10.02\\                       
\noalign{\smallskip}\bottomrule
%$\geq$ 30 TeV,
\end{tabular}
\end{center}
\end{table}

%%%%%%%%%%%%%%%%%%%%%%%%
% Section : Analysis %%%
%%%%%%%%%%%%%%%%%%%%%%%%
\section{Air Shower Data Analysis}
\label{sec:shower-analysis}

\subsection{Preprocessing of Simulated Data}
\label{sec:preprosession}

The energy deposited in each AS detector is converted to the parameter $\rho$, defined as the number of minimum ionizing particles (MIPs) per square meter.
Here, the energy deposited by a vertically incident MIP is taken to be 5.6\,MeV.
AS detectors recording $\rho > 3.5$\,m$^{-2}$ are designated as {\it hit} detectors.
The core position, arrival direction, and total $\rho$ ($\Sigma \rho$) of each shower event are reconstructed using the standard analysis procedure employed in the Tibet experiment \cite{Amenomori2009Moonshadow}.

The output of the MD simulation corresponds to the number of photoelectrons detected by the PMT in each MD cell.
This number is normalized by the average number of photoelectrons (typically 15\,p.e.) produced by a muon passing through a single MD cell and is stored as the parameter $N_{\mu}$.
The total number of muons for each shower, $\Sigma N_{\mu}$, is then obtained by summing over all MD cells.

\subsection{Selection of Well-Reconstructed Shower Events}
\label{sec:event-selection}

To select well-reconstructed air-shower events from the simulated dataset, the following selection criteria were applied to the AS-array data.
In addition to the standard criteria used in the Tibet experiment, stricter requirements were imposed to ensure that the shower core falls within the coverage area of the MD array:

\begin{enumerate}
\item At least four {\it hit} detectors must register signals within a 600\,ns time window.
\item At least four {\it hit} detectors must be located within the inner detector region, defined as those not on the outermost boundary of the AS array (see Fig.~\ref{fig:Tibet-III+MD}).
\item The three detectors recording the highest $\rho$ values must be located within 80\,m of the array center. This 80\,m radius region is indicated by the gray hatched area in Fig.~\ref{fig:Tibet-III+MD}.
\item The reconstructed shower core position must lie within a 70\,m radius from the array center. This defines the geometrical acceptance area, $S_\textrm{geom} = 15{,}400$\,m$^2$, as shown by the solid circle in Fig.~\ref{fig:Tibet-III+MD}.
\item The sum of the residuals between the measured particle arrival times and the best-fit shower front (assumed to have a reverse-conical shape \cite{Tibet2003}) must be less than 1.0\,m, assuming that the particles travel at the speed of light.
\item The reconstructed zenith angle $\theta$ must satisfy $\sec \theta < 1.1$, corresponding to a geometrical solid angle of $\Omega_\textrm{geom} = 0.571$\,sr.
\end{enumerate}

\subsection{Selection Based on MD Data}
\label{sec:md-selection}

\begin{figure}[htbp]
	\begin{center}
		\includegraphics[width=0.7\linewidth]{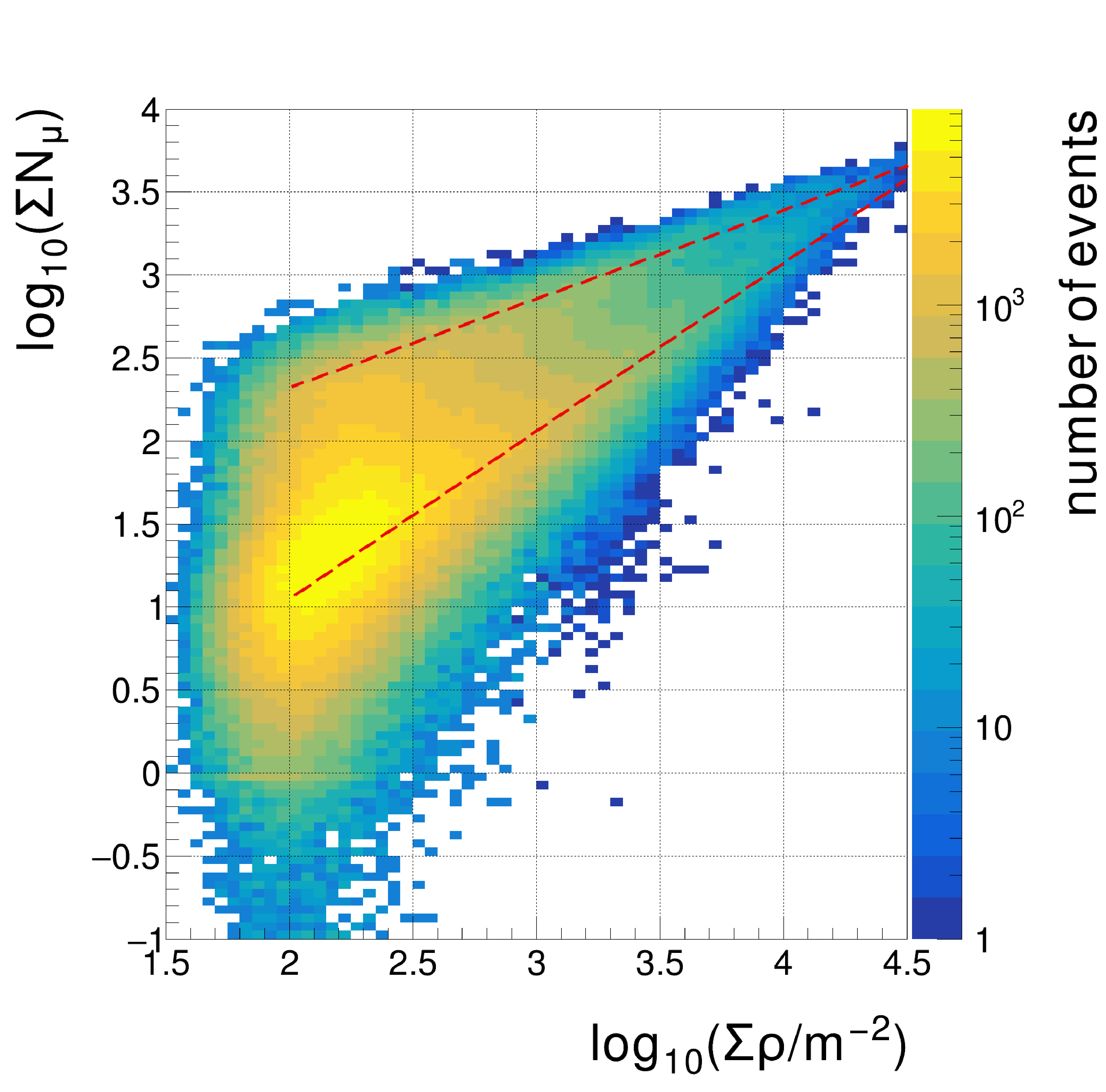}
	\end{center}
		
        \caption{Two-dimensional distribution of $\Sigma \rho$ and $\Sigma N_{\mu}$ from the MC dataset generated using the EPOS-LHC/Shibata model combination. The color scale indicates the number of reconstructed air-shower events in each bin. The red line above corresponds to events where the shower core falls directly over an MD cell, while the red line below corresponds to events where the shower core falls away from any MD cell.}

 		\label{fig:sumrho-summu-original}
\end{figure}

\noindent
The two-dimensional distribution of the reconstructed parameters $\Sigma \rho$ and $\Sigma N_{\mu}$ forms the basis of the unfolding analysis described in Sec.~\ref{sec:unfolding}.
Figure~\ref{fig:sumrho-summu-original} presents the distribution obtained from a Monte Carlo dataset generated using the EPOS-LHC and Shibata models.
Since $\Sigma \rho$ is approximately proportional to the energy of the primary particle, the figure demonstrates that $\Sigma N_{\mu}$ increases with increasing energy.

The spread in the $\Sigma N_{\mu}$ distribution reflects the variation in mass composition at each energy.
However, as seen in Fig.~\ref{fig:sumrho-summu-original}, the correlation between $\Sigma \rho$ and $\Sigma N_{\mu}$ exhibits a bifurcation into two distinct trends above and below $\log_{10}(\Sigma N_{\mu}) \approx 2$.
As will be discussed below, this feature is not attributed to differences in the primary cosmic-ray composition.

%%% Fig.-1 %%%%%%%%%%%%%%%%%%%%%%%%%%%%%%%%%%%%%%%%%%%%%%%%%

The double trends appear depending on the position of the shower core.  
When a shower core falls above an MD cell, $\Sigma N_{\mu}$ tends to show a larger value because of the presence of high-energy penetrating particles around the shower core.  
In addition, since the number density of muons rapidly drops with distance from the shower core, $\Sigma N_{\mu}$ becomes very sensitive when the shower core is located near the area covered by the MDs.  
Inclusion of the MD cells located near the shower core results in a strong position dependence.  
In contrast, as seen in Fig.~\ref{fig:Tibet-III+MD}, when a shower falls at the center of the array, the minimum distance to the MD cells is 30\,m, which systematically reduces $\Sigma N_{\mu}$.

To reduce the strong dependence of the shower core position on the MD arrangement, we redefine $\Sigma N_{\mu}$ by introducing a minimum distance from the shower core position to the MD cells to be integrated, as

\begin{equation}
\label{eq:totalmuon}
(\Sigma N_{\mu})_{31\textrm{m}} = \sum^{\geq 31\,\textrm{m}} N_{\mu} \times \left(\frac{64}{64 - N^{<31\,\textrm{m}}}\right),
\end{equation}

where $\sum^{\geq 31\,\textrm{m}} N_{\mu}$ is the number of muons determined from the MD cells located at a distance of 31\,m or more from the shower core, and $N^{<31\,\textrm{m}}$ is the number of MD cells within 31\,m of the shower core.  
The value is normalized to the original definition in which all 64 MD cells are taken into account.  
The distribution of $\Sigma \rho$ and $(\Sigma N_{\mu})_{31m}$ with this new definition, shown in Fig.~\ref{fig:sumrho-summu-mod}, no longer exhibits the double trends seen in Fig.~\ref{fig:sumrho-summu-original}.

\subsection{Basic Properties of the $\Sigma \rho$--$\Sigma N_{\mu}$ Distribution}
\label{sec:properties}

\begin{figure}[htbp]
\begin{center}
\includegraphics[width=0.7\linewidth]{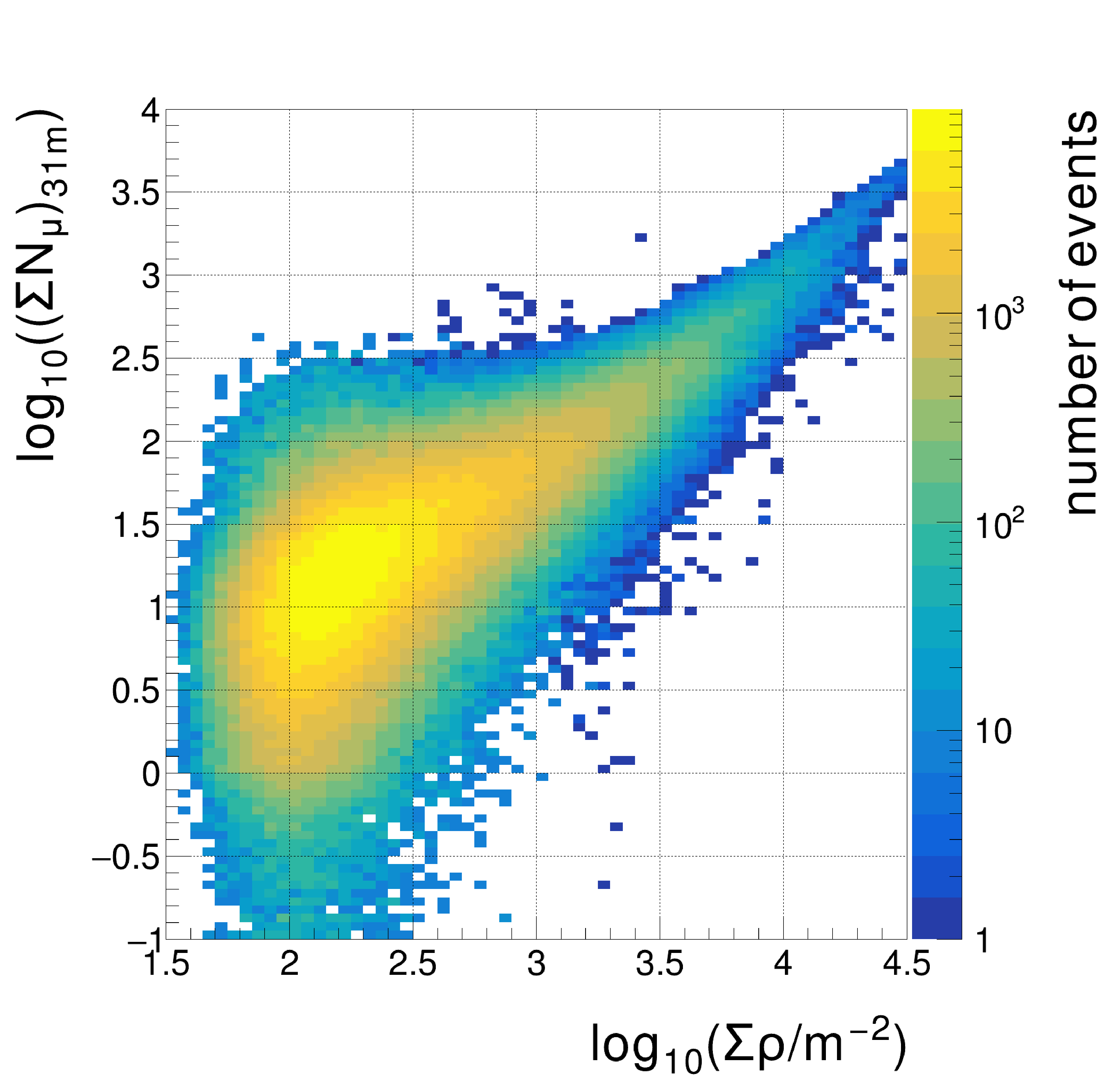}
\end{center}
\caption{Same as Fig.\ref{fig:sumrho-summu-original} but using the new definition of $(\Sigma N_{\mu})_{31m}$ given in Eq.\ref{eq:totalmuon}.
}.
\label{fig:sumrho-summu-mod}
\end{figure}

\begin{figure}[htbp]
\begin{center}
 \includegraphics[width=0.7\linewidth]{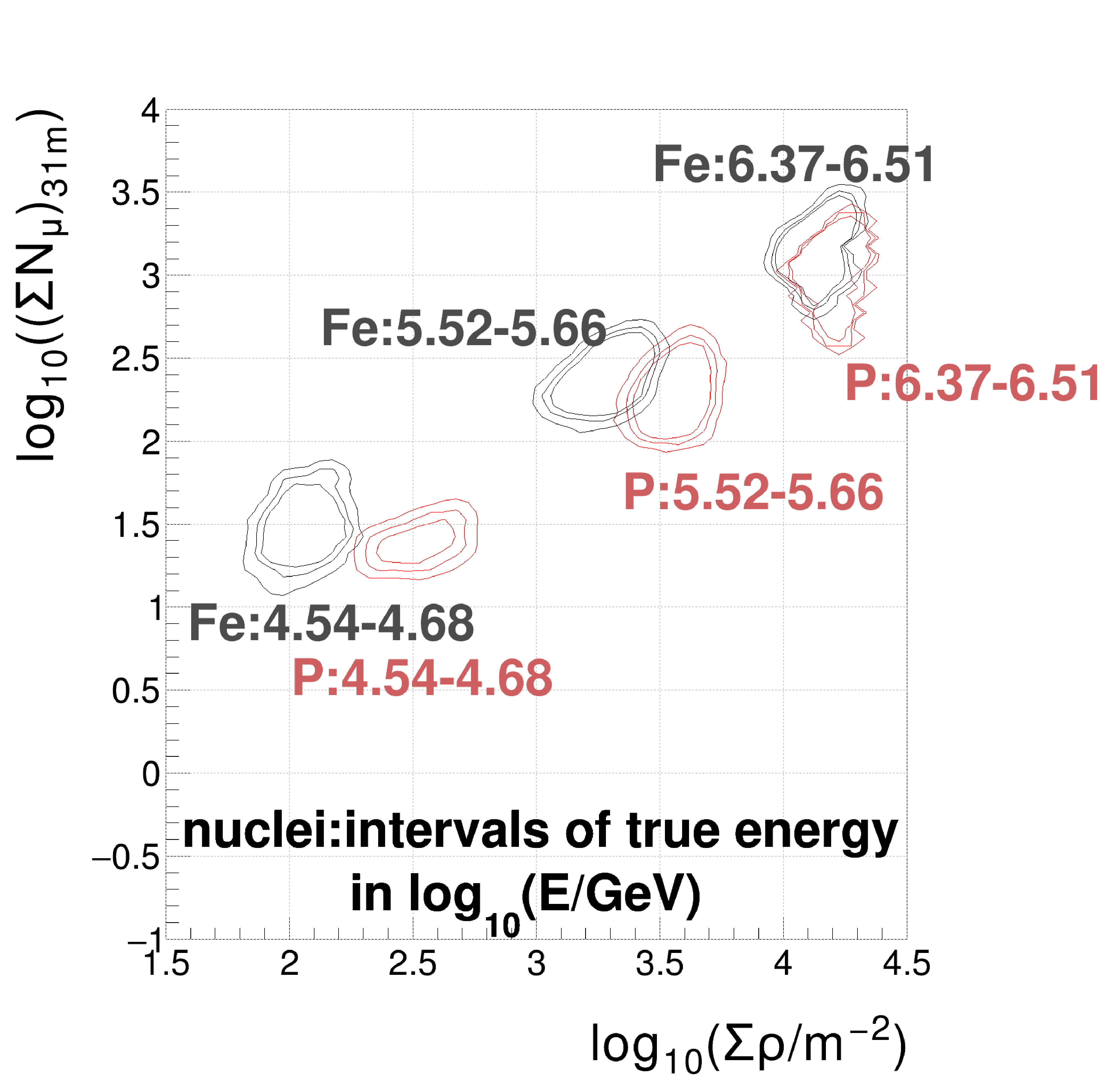}
\end{center}
\caption{
Distributions of $\Sigma \rho$ and $(\Sigma N_{\mu})_{31m}$ for P (red) and Fe (black) primaries in narrow energy bins: $4.54 < \textrm{log}_{10}(E/\textrm{GeV}) < 4.68$, $5.52 < \textrm{log}_{10}(E/\textrm{GeV}) < 5.66$, and $6.37 < \textrm{log}_{10}(E/\textrm{GeV}) < 6.51$.  
The isolines, from innermost to outermost, represent probability densities of 0.005, 0.003, and 0.001.  
For example, the bins between the outermost and the next inner contour lines contain 0.1\% to 0.3\% of the events within the same energy band and the same mass group.
}

\label{fig:sumrho-summu-contour}
\end{figure}

\begin{figure}[htbp]
\centering

\begin{minipage}[b]{0.60\linewidth}
\centering
\includegraphics[width=1.0\linewidth]{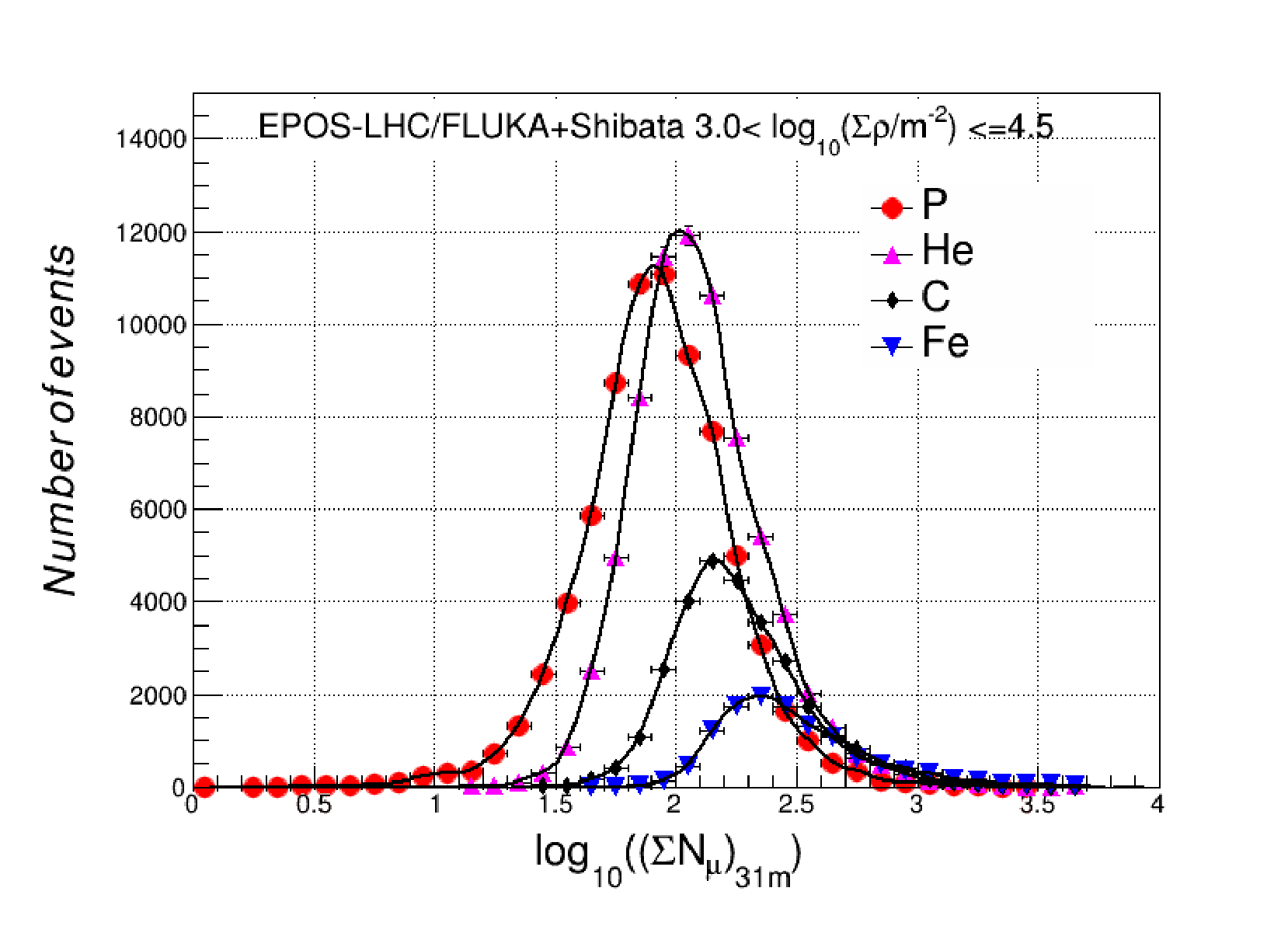}
\subcaption{}
\end{minipage}
%\hspace{0.05\linewidth} % Space between images

%\vspace{0.05\linewidth} % Space between rows

\begin{minipage}[b]{0.60\linewidth}
\centering
\includegraphics[width=1.0\linewidth]{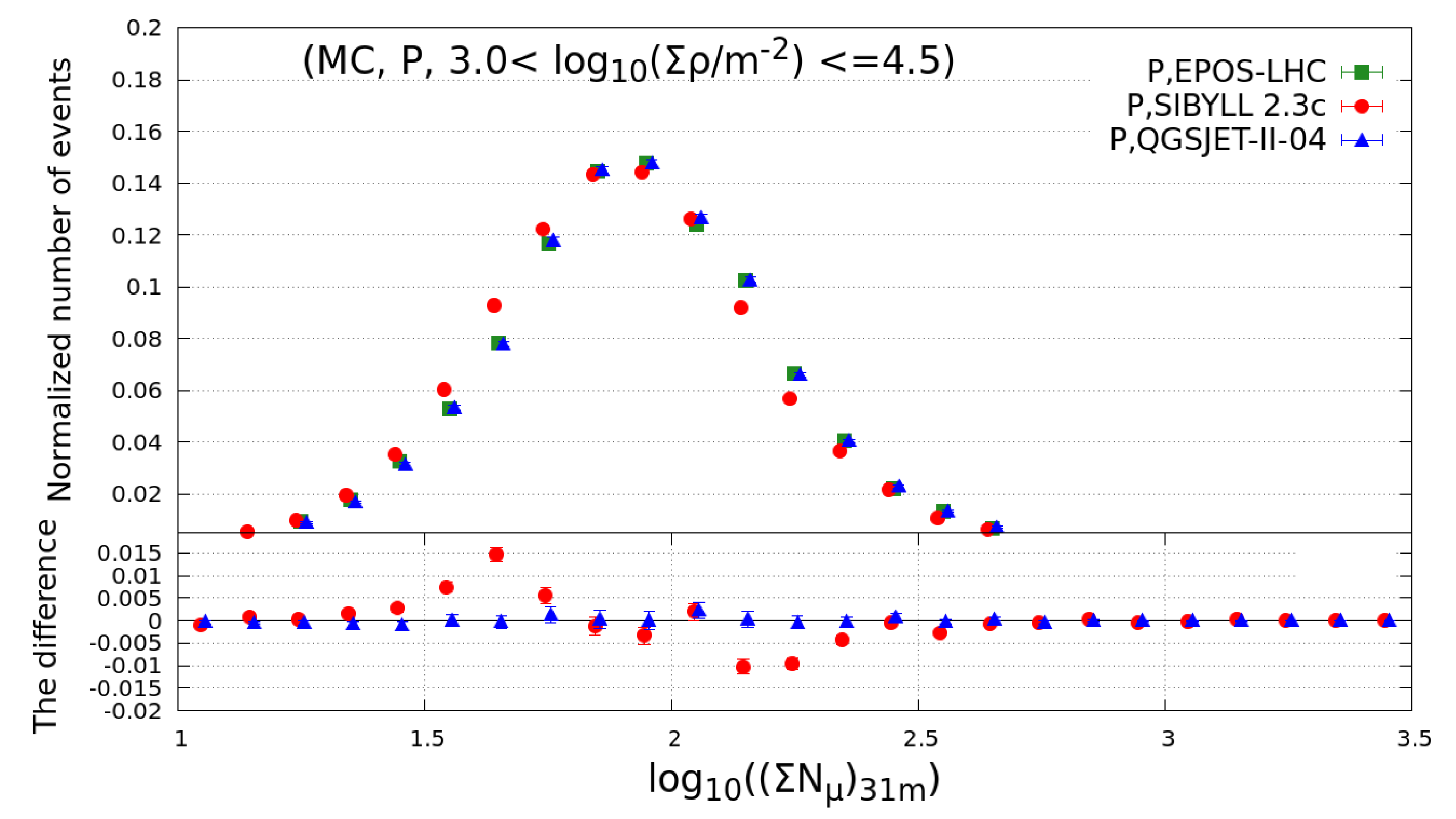}
\subcaption{}
\end{minipage}
%\hspace{0.05\linewidth} % Space between images
\begin{minipage}[b]{0.60\linewidth}
\centering
\includegraphics[width=1.0\linewidth]{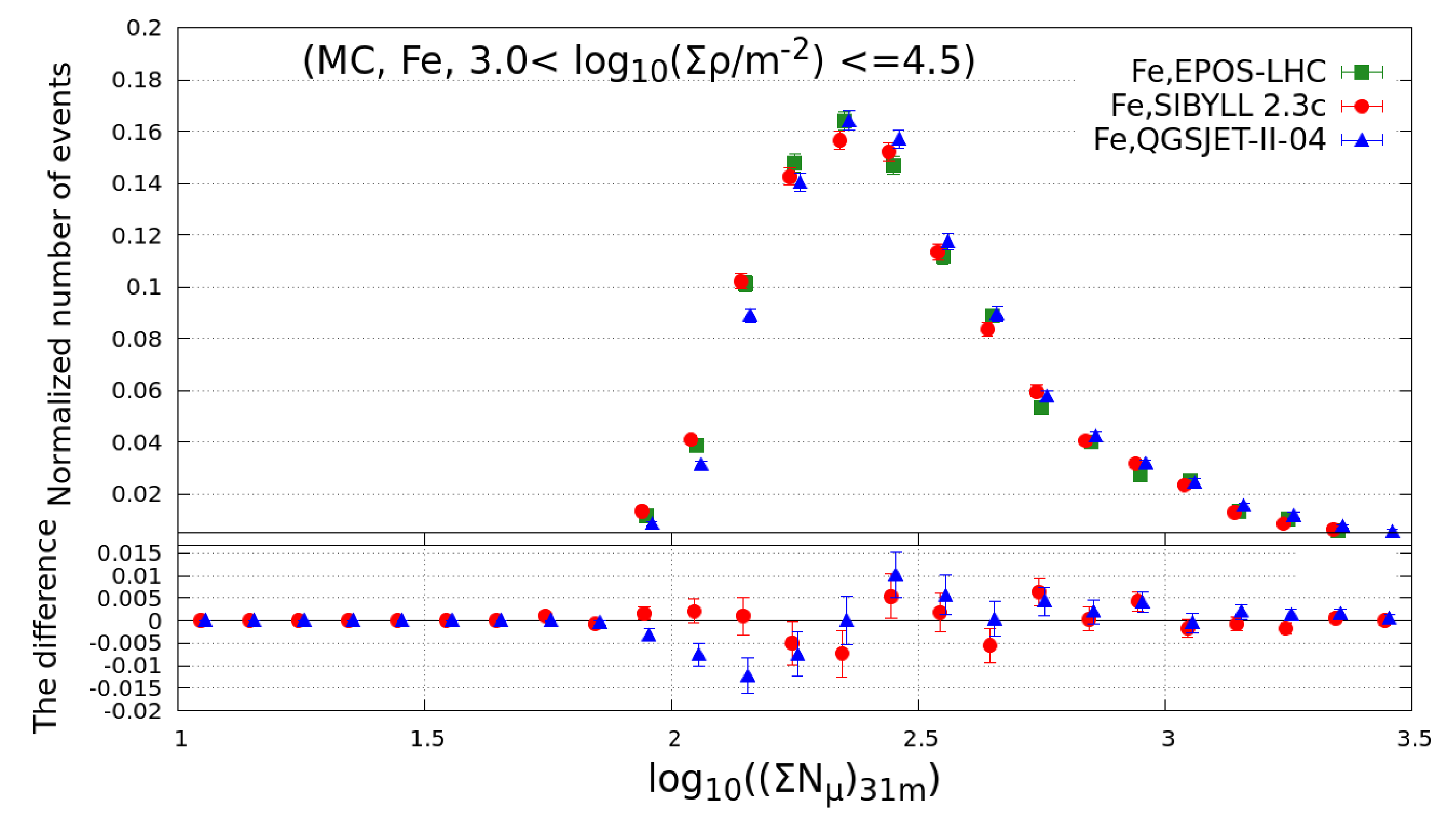}
\subcaption{}
\end{minipage}

\caption{
Distributions of $(\Sigma N_{\mu})_{31\textrm{m}}$ for events with $3.0 < \textrm{log}_{10}(\Sigma\rho/\textrm{m}^{-2}) < 4.5$.  
(a) Comparison of primary particles: P (red), He (magenta), C (black), and Fe (blue).  
(b) and (c) show the differences between high-energy interaction models. The upper panels display normalized event numbers: green, red, and blue points represent EPOS-LHC, SIBYLL 2.3c and QGSJET-II-04, respectively. The lower panels show the ratios to the EPOS-LHC results. The points for SIBYLL 2.3c and QGSJET-II-04 are slightly shifted for visibility.
}

\label{fig:1dslices}
\end{figure}

\noindent  
The $\Sigma \rho$--$(\Sigma N_{\mu})_{31m}$ distribution shown in Fig.~\ref{fig:sumrho-summu-mod} is a convolution of the continuous energy spectra of several mass groups.  
While unfolding this convoluted distribution using the two-dimensional response matrix is the aim of this study, we first demonstrate  
some basic distributions for specific energy ranges and mass groups.  
Figure~\ref{fig:sumrho-summu-contour} shows the $\Sigma \rho$--$(\Sigma N_{\mu})_{31\textrm{m}}$ distributions for proton and iron primaries in the energy ranges of 10$^{4.54-4.69}$\,GeV, 10$^{5.52-5.66}$\,GeV, and 10$^{6.37-6.51}$\,GeV.  
It is seen that the separation between the two groups is clear at lower energies but gradually diminishes at higher energies.  
Figure~\ref{fig:1dslices}(a) shows one-dimensional slices of $(\Sigma N_{\mu})_{31\textrm{m}}$ in the range 3.0 $<$ log$_{10}(\Sigma \rho$/m$^{-2}) \leq$ 4.5 for P, He, C, and Fe.  
Although the distributions overlap, the peak positions clearly depend on the mass group.  
Figures~\ref{fig:1dslices}(b) and (c) compare the $(\Sigma N_{\mu})_{31\textrm{m}}$ distributions among three different hadronic interaction models in the same $\Sigma \rho$ range for P and Fe, respectively.  
These differences are unavoidable and constitute a major source of systematic uncertainty, to be discussed in Sec.~\ref{sec:results}.

While various methods such as machine learning and unfolding have been applied to extract the energy spectra of individual mass groups from this relationship \cite{Xishui Tian,KASCADE-GrandeD,Pierre Auger Collaboration)},  
we examine the Bayesian unfolding method in this study.

%%% Fig.-6 %%%%%%%%%%%%%%%%%%%%%%%%%%%%%%%%%%%%%%%%%%%%%%%%%

%%% Fig.-6 %%%%%%%%%%%%%%%%%%%%%%%%%%%%%%%%%%%%%%%%%%%%%%%%%

%%%%%%%%%%%%%%%%%%%%%%%%%
% Section : Unfolding %%%
%%%%%%%%%%%%%%%%%%%%%%%%%
\section{Unfolding Method}
\label{sec:unfolding}

\noindent
The number of events in the $i$-th bin $N_i$ of the $\Sigma \rho$--$(\Sigma N_{\mu})_{31\textrm{m}}$ distribution is related to the number of events in the $j$-th primary energy bin of mass element $k$ denoted by $x^k_j$ through the response matrix $R^k_{ij}$ the acceptance $A^k_j$ (defined as the product of effective area and solid angle including detection efficiency) and the observation time $T$ as

\begin{equation} \label{eq:defrelation}
    N_i(\Sigma \rho, (\Sigma N_{\mu})_{31\textrm{m}}) = T \sum_{k=1}^{N_{\text{nucl}}} \sum_{j=1}^{N_s} R^k_{ij} \, A^k_j \, x^k_j.
\end{equation}

Here, $N_{\text{nucl}}$ and $N_s$ are total numbers of mass groups and primary energy bins, respectively.  
The element $x^k_j$ is defined as

\[
x^k_j = \displaystyle\int\limits_{E_j}^{E_j + \Delta E_j}\cfrac{dJ_k}{dE} \, dE,
\]

where $\Delta E_j$ is the energy bin width.  
The differential flux of element $k$, $dJ_k/dE$ is the quantity we aim to derive.  
The response matrix is constructed as

\begin{equation} \label{eq:defresponse}
R^k_{ij} 
= \cfrac{\displaystyle\int\limits_{E_j}^{E_j+\Delta E_j}
            p_k(i|E) \, A^k_j \, \cfrac{dJ^\textrm{model}_k}{dE} \, dE}
        {\displaystyle\int\limits_{E_j}^{E_j+\Delta E_j}
             A^k_j \, \cfrac{dJ^\textrm{model}_k}{dE} \, dE},
\end{equation}

where $p_k(i|E)$ gives the probability that a detected cosmic ray of energy $E$ and mass group $k$ is observed in the $i$-th bin.  
This probability depends on the hadronic interaction model and is normalized as $\sum_i p_k(i|E) = 1$.

In practice, $R^k_{ij}$ also depends on the assumed composition model $dJ^{\textrm{model}}_k/dE$.  
When the energy bin width $\Delta E_j$ is sufficiently narrow such that $p_k(i|E)$ can be considered energy-independent within the bin, $R^k_{ij}$ becomes identical to $p_k(i|E_j)$ and independent of the composition model.  
Due to the finite Monte Carlo statistics in this study, we use the definition in Eq.~\ref{eq:defresponse}, meaning that $R^k_{ij}$ may exhibit some dependence on the composition model.

In this paper, the response matrix $R^k_{ij}$ is calculated using a Monte Carlo dataset based on the EPOS-LHC hadronic interaction model and the Shibata composition model.  
The test air-shower datasets are constructed using other model combinations.  
The dataset based on EPOS-LHC and the Gaisser composition model is also analyzed to evaluate performance independent of the choice of interaction model.

\subsection{Effective Acceptance of Air Shower Detection}
\label{sec:acceptance}

\noindent
The effective acceptance $A^k_j$ introduced in Eq.~\ref{eq:defrelation} is defined as:

\begin{equation}
A^k_j = \frac{N^k_{j,\textrm{survived}}}{N^k_j} \times S_\textrm{sim} \times \Omega_\textrm{sim},
\end{equation}

where $N^k_j$ and $N^k_{j,\textrm{survived}}$ are the numbers of events of mass element $k$ in the $j$-th energy bin that are generated and survived after the selection conditions described in Sec.~\ref{sec:event-selection}, respectively.  
Here, $S_{\textrm{sim}}$ and $\Omega_\textrm{sim}$ correspond to the area within a 300\,m radius from the array center and the solid angle for zenith angles below $45^{\circ}$ as mentioned in Sec.~\ref{sec:shower-analysis} and Tab.~\ref{tab:1}.

\begin{figure}[H]
	\begin{center}
		\includegraphics[width=0.7\linewidth]{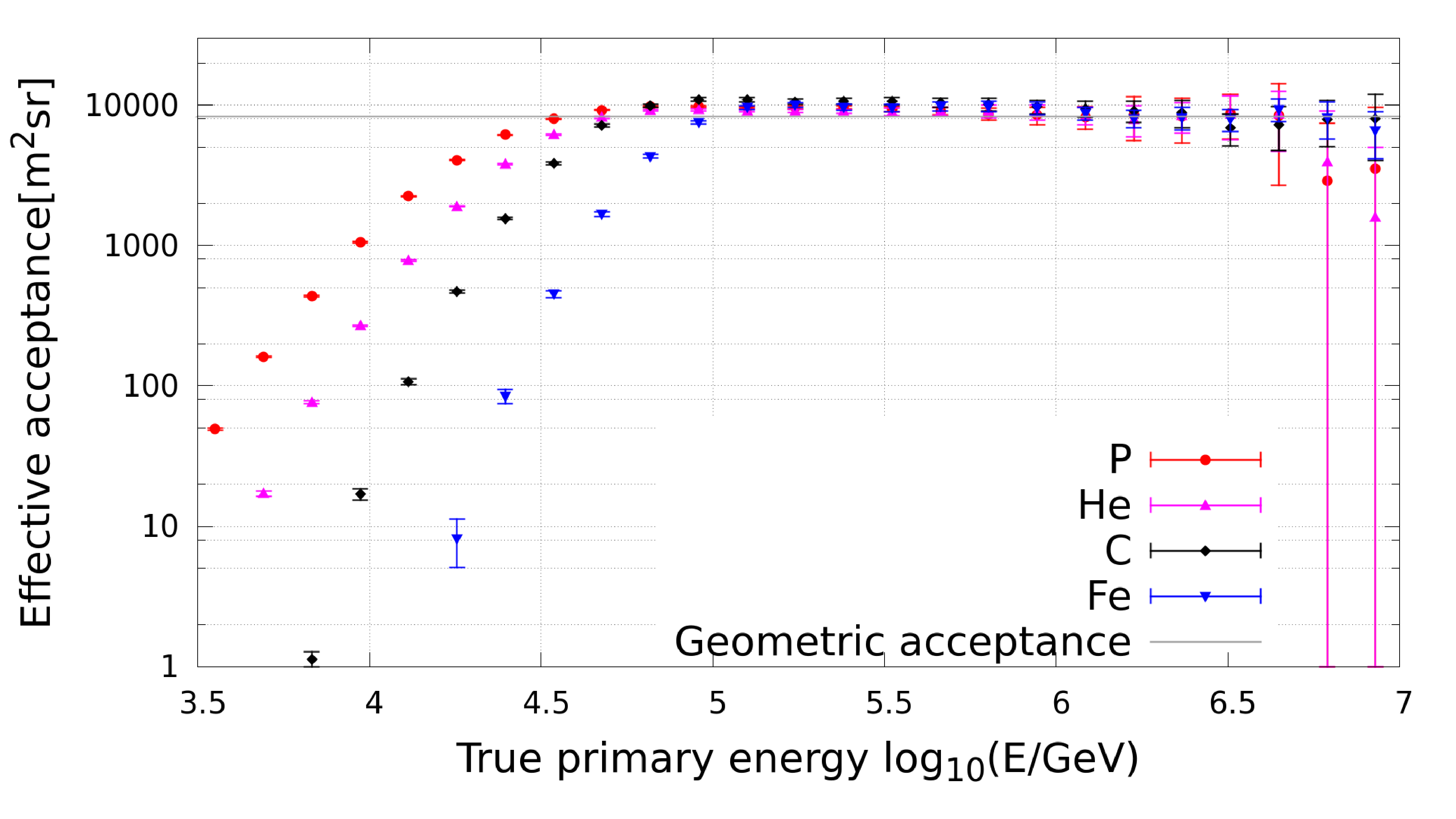}
	\end{center}
		\caption{
Effective acceptances for P (red), He (magenta), C (black), and Fe (blue) primaries.
The horizontal line indicates the geometrical acceptance.
}
 		\label{fig:acceptance}
\end{figure}

The acceptances calculated for the proton, helium, carbon, and iron primaries are shown in Fig.~\ref{fig:acceptance}.  
For all nuclei, the acceptances are nearly constant above $10^{5}$\,GeV.  
While the geometrical acceptance defined in Sec.~\ref{sec:event-selection} is $S_\textrm{geom} \times \Omega_\textrm{geom} = 8796$\,m$^2$\,sr, the effective acceptances exceed this value due to misreconstructed events that migrate from outside the defined core or zenith angle boundaries.

In the following analysis, we use data above an energy threshold specific to each nuclear type to ensure constant acceptance:  
$10^{4.5}$\,GeV for protons and helium, $10^{4.8}$\,GeV for the carbon group, and $10^{5.1}$\,GeV for the iron group.  
The acceptance $A^k_j$ is calculated using the same Monte Carlo dataset employed for the response matrix—based on the EPOS-LHC hadronic interaction model and the Shibata composition model.

\subsection{Reconstruction of CR Spectra}

The simulated event datasets $N_i(\Sigma \rho, (\Sigma N_{\mu})_{31\textrm{m}})$ in Eq.~\ref{eq:defrelation} are unfolded to $x^k_j$.  
There are various methods to solve equations like Eq.~\ref{eq:defrelation}.  
Here, 
We use the iterative Bayesian unfolding method \cite{G. D’Agostini}, which is known to provide more stable results than other unfolding techniques, with reduced sensitivity to statistical fluctuations. In this approach, the detector response, obtained from MC, is used to relate the true and measured distributions. Since the exact inverse of the response function does not always exist, an effective inverse is constructed iteratively using Bayes’ theorem. Starting from an assumed prior distribution, successive updates are performed until the result becomes nearly unchanged. To further control statistical fluctuations, a smoothing procedure was applied after each iteration \cite{Igor Volobouev, Manfred Matzke}. The detailed procedure of the algorithm is described, for example, in Ref.~\cite{KASCADE-GrandeD}.

%We use the iterative Bayesian unfolding method \cite{G. D’Agostini}, which is known to provide stable results with reduced sensitivity to statistical fluctuations. In this analysis, the response matrix was constructed from Monte Carlo simulations. To control statistical fluctuations, a smoothing procedure was applied after each iteration \cite{Igor Volobouev,Manfred Matzke}, and the unfolding was repeated until the solution became nearly unchanged, ensuring stable and reliable results. The detailed procedure of the Bayesian unfolding algorithm is described, for example, in Ref.~\cite{KASCADE-GrandeD}.

The differential flux is then determined as

\begin{equation}
\label{Eq:flux1}
\frac{dJ_k}{dE} = \frac{x^k_j}{T\Delta E_j},
\end{equation}

where the observation time $T$ corresponding to the simulated events is determined from the number of simulated events, $S_{\textrm{sim}}$, $\Omega_{\textrm{sim}}$ and the Shibata flux model, resulting in $T = 229{,}209$\,sec, which is equivalent to 2.65\,days of observation.

%%%%%%%%%%%%%%%%%%%%%%%%%
%%% Section : Results %%%
%%%%%%%%%%%%%%%%%%%%%%%%%
\section{Results}
\label{sec:results}
We applied the unfolding analysis to datasets produced using the EPOS-LHC/Gaisser, SIBYLL 2.3c/Gaisser and QGSJET~II-04/Gaisser model combinations.  
Note that the response matrix $R^k_{ij}$ and the acceptance $A^k_j$ are constructed from the dataset generated using the EPOS-LHC/Shibata model combination.  
Therefore, the unfolding result for the EPOS-LHC/Gaisser dataset reflects uncertainties due to the unfolding technique and the composition model assumption used in constructing the response matrix and acceptance, while remaining free from uncertainties related to the hadronic interaction model.

To estimate statistical uncertainties, instead of generating independent datasets—which would require substantial computational resources—pseudo datasets were created by fluctuating the event counts $N_i(\Sigma \rho, (\Sigma N_{\mu})_{31\textrm{m}})$ according to the Poisson distribution.  
These pseudo datasets were analyzed in the same way as the original Monte Carlo datasets.

\begin{figure}[H]
	\begin{center}
		\includegraphics[width=0.9\linewidth]{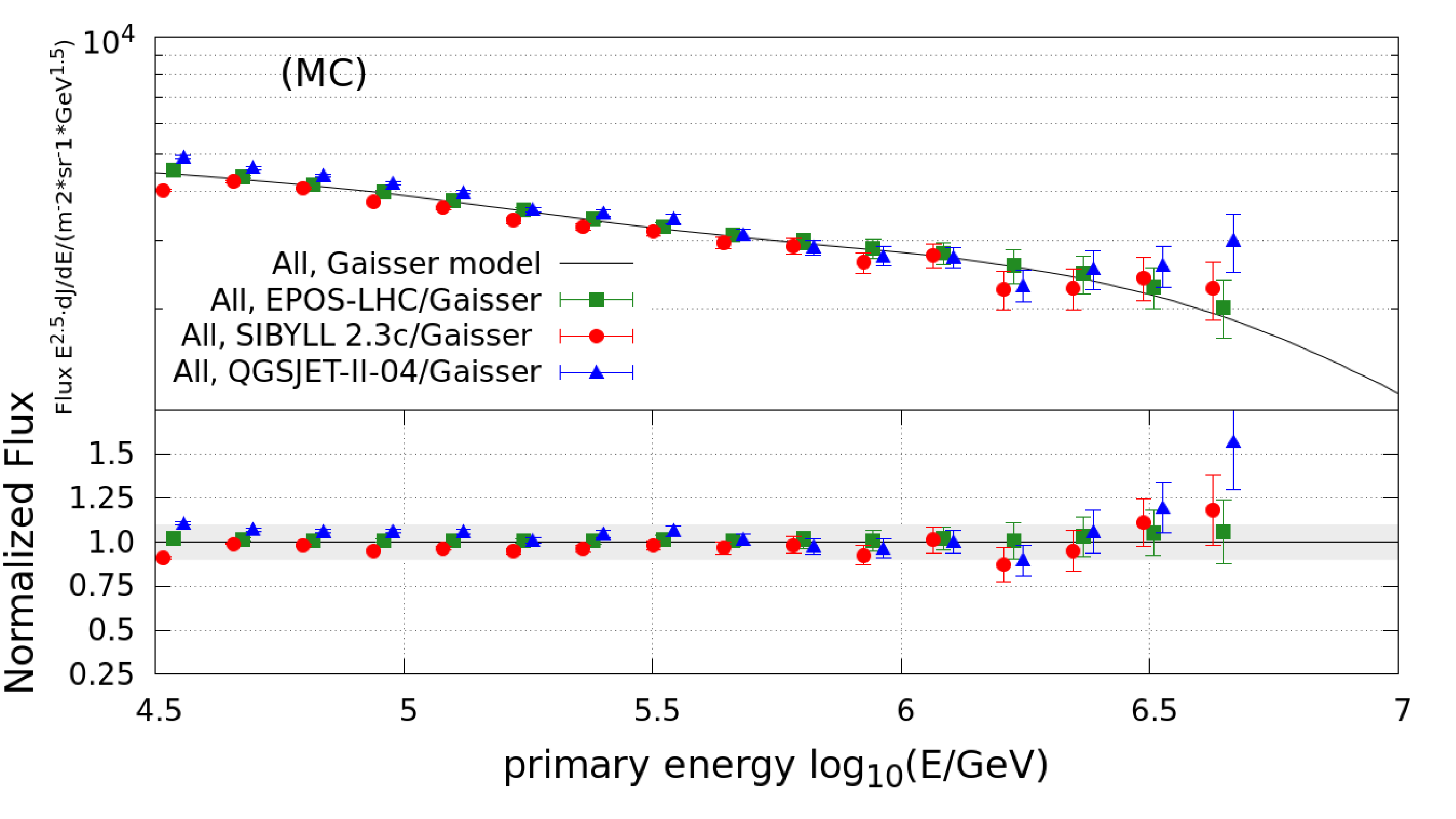}
	\end{center}
\caption{
Comparison of the unfolded all-particle energy spectra.  
Green, red, and blue points show the results for datasets produced using the EPOS-LHC/Gaisser, SIBYLL 2.3c/Gaisser, and QGSJET~II-04/Gaisser model combinations, respectively.  
The black solid curve indicates the input spectrum of the Gaisser model.  
The bottom panel shows the ratio to the model flux.  
The gray hatched area indicates a $\pm$10\% variation from the model flux.  
The SIBYLL 2.3c and QGSJET~II-04 points are slightly shifted laterally from the EPOS points for visibility.
}

 		\label{fig:All_spectrum}
\end{figure}

The unfolded all-particle spectra for the three datasets are shown in Fig.~\ref{fig:All_spectrum}.  
The input spectrum based on the Gaisser model is shown as a black line, and the ratio to the input flux is presented in the bottom panel.  
Up to $10^{6.5}$\,GeV, the difference from the input flux remains within $\pm$10\%, as indicated by the gray hatched area.  
At $10^{6.7}$\,GeV, the maximum deviation reaches +58\%.  
Since the deviations above $10^{6}$\,GeV are comparable to the statistical errors, the performance in this region may be limited by the Monte Carlo statistics (i.e., the equivalent observation time of 2.65 days).  
The green points, which represent the EPOS-LHC/Gaisser dataset unfolded using the EPOS-LHC/Shibata response matrix and acceptance, show deviations within $\pm$10\% across the entire energy range.

\begin{figure}[H]
	\begin{center}
		\includegraphics[width=0.9\linewidth]{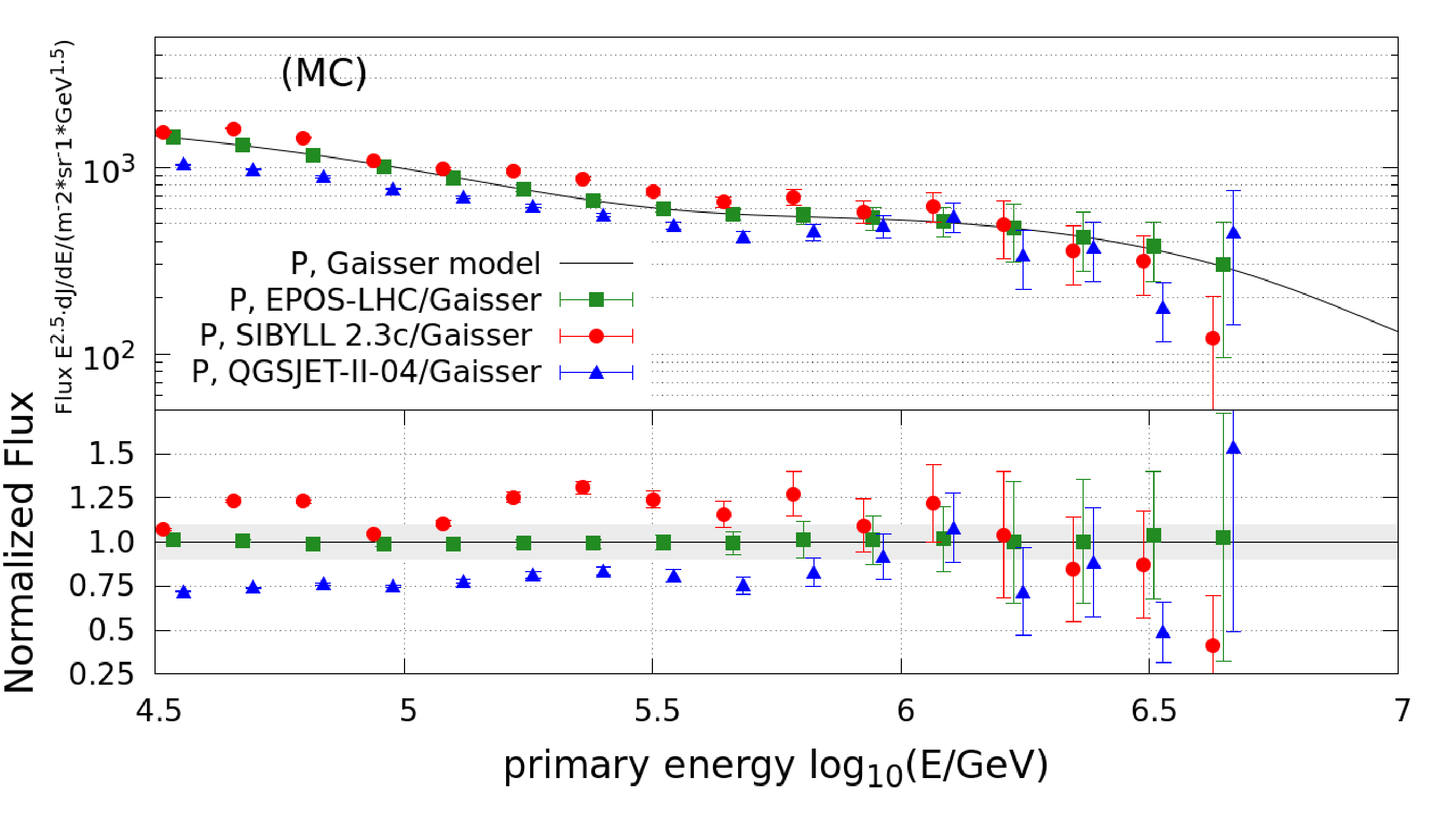}
	\end{center}
		\caption{Comparison of the unfolded proton energy spectra. Same as Fig.\ref{fig:All_spectrum}.}
 		\label{fig:P_spectrum}
\end{figure}

\begin{figure}[htbp]
	\begin{center}
		\includegraphics[width=0.9\linewidth]{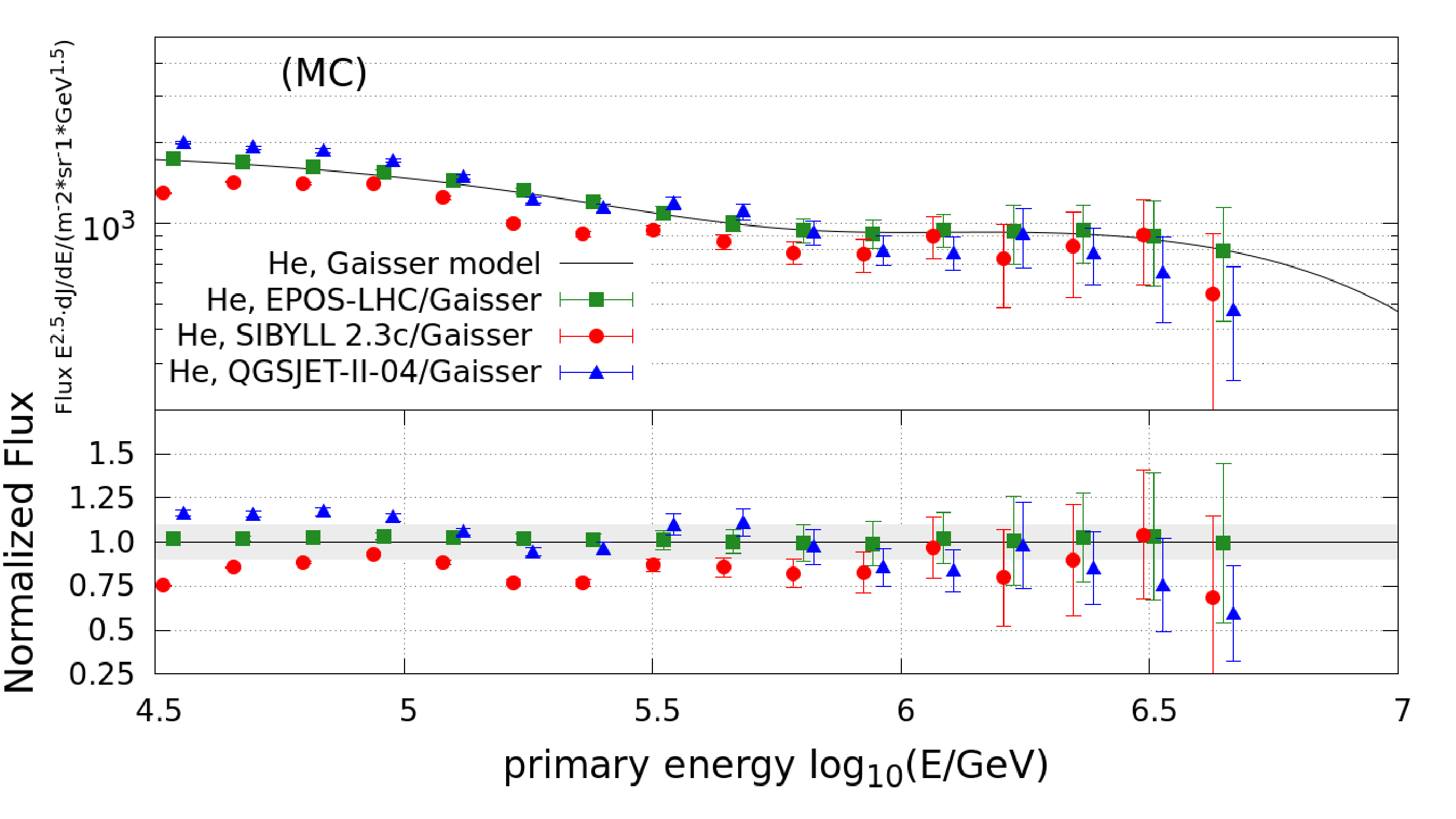}
	\end{center}
		\caption{Comparison of the unfolded Helium energy spectra. Same as Fig.\ref{fig:All_spectrum}.}
 		\label{fig:He_spectrum}
\end{figure}

\begin{figure}[H]
	\begin{center}
		\includegraphics[width=0.9\linewidth]{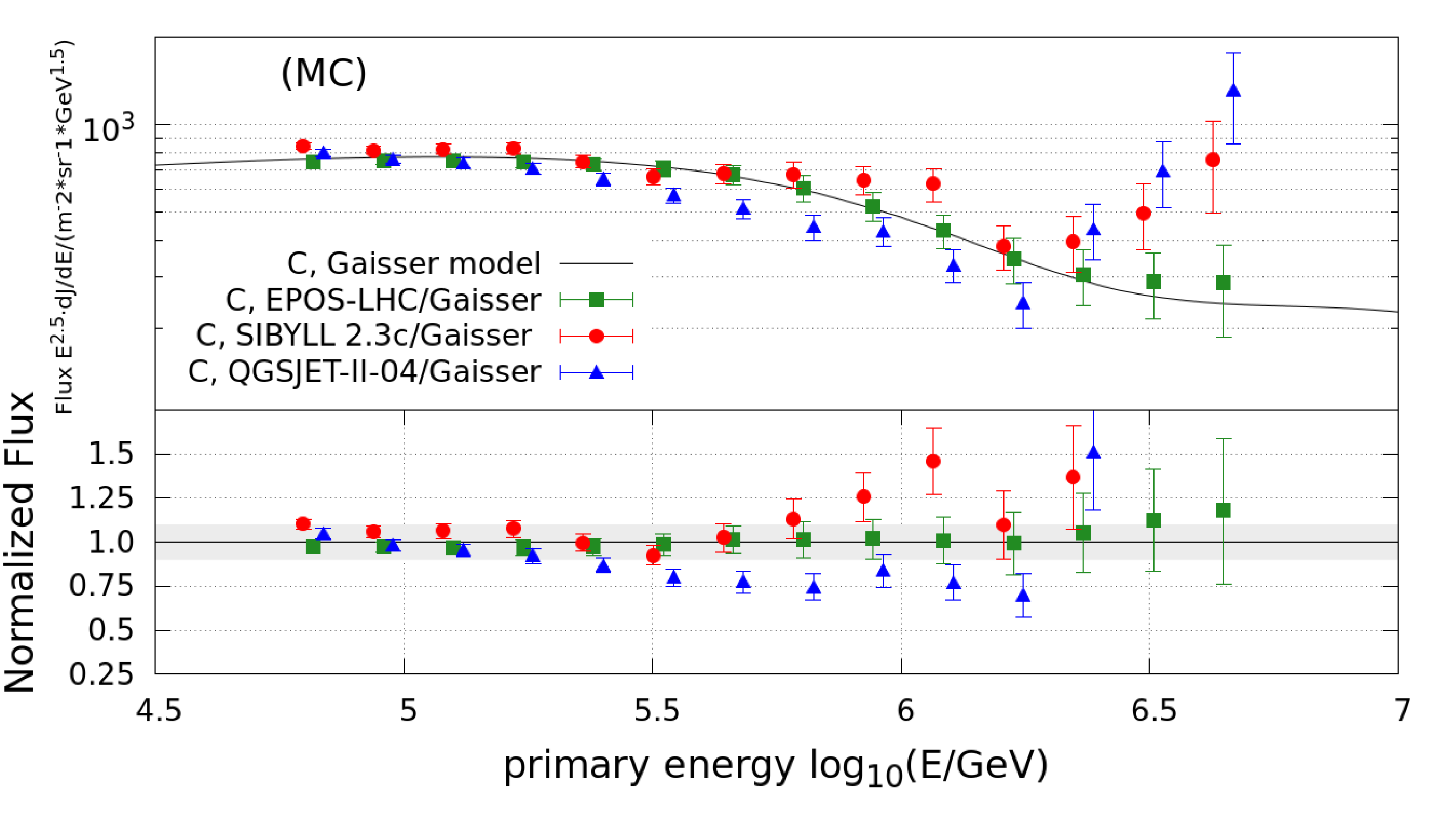}
	\end{center}
		\caption{Comparison of the unfolded Carbon energy spectra. Same as Fig.\ref{fig:All_spectrum}.}
 		\label{fig:C_spectrum}
\end{figure}

\begin{figure}[H]
	\begin{center}
		\includegraphics[width=0.9\linewidth]{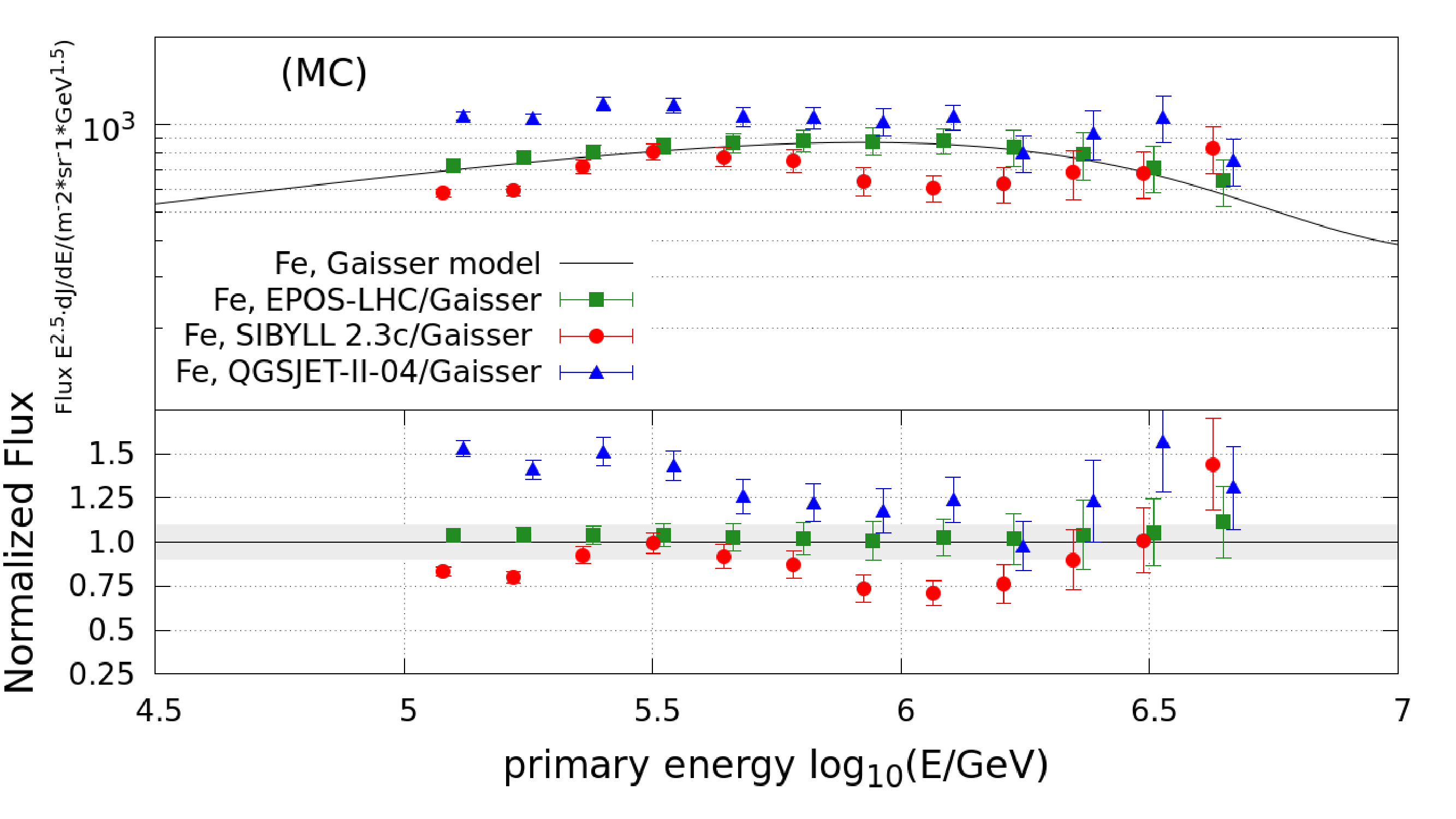}
	\end{center}
		\caption{Comparison of the unfolded Iron energy spectra. Same as Fig.\ref{fig:All_spectrum}.}
 		\label{fig:Fe_spectrum}
\end{figure}

The unfolded spectra of individual mass groups are shown in Fig.~\ref{fig:P_spectrum}, Fig.~\ref{fig:He_spectrum}, Fig.~\ref{fig:C_spectrum}, and Fig.~\ref{fig:Fe_spectrum} for P He, C, and Fe, respectively.  
In all cases, the deviations of the green points from the input flux are within $\pm$10\%, except at a few points in the highest energy bins.  
In contrast, the red and blue points exhibit larger deviations, indicating that the dominant uncertainty in this analysis arises from the choice of hadronic interaction model.

The model dependence of the proton and helium flux reconstructions is within $\pm$25\% below $10^{6.5}$\,GeV.  
For carbon, the model dependence is also within $\pm$25\% below $10^{6}$\,GeV but it increases significantly above $10^{6.5}$\,GeV.  
For iron, a large model dependence is observed even at lower energies, reaching +55\% to -30\% over the energy range of $10^{5.1}$\,GeV to $10^{6.7}$\,GeV.

%%% Fig.-10 %%%%%%%%%%%%%%%%%%%%%%%%%%%%%%%%%%%%%%%%%%%%%%%%%


%%%%%%%%%%%%%%%%%%%%%%%%%%%%%
%%% Section : Conclusions %%%
%%%%%%%%%%%%%%%%%%%%%%%%%%%%%
\section{Conclusion}
\label{sec-conclusion}

We studied a reconstruction method for the cosmic-ray energy spectra of individual mass groups, assuming the use of data taken by the Tibet-AS and MD arrays in the energy range from $10^{4.5}$\,GeV to $10^{6.5}$\,GeV.  
This energy range is important for understanding the origin and propagation of cosmic rays in our galaxy.  
The large underground muon detectors and the high operational altitude of the Tibet experiment offer a unique sensitivity for such studies.  
We applied the iterative Bayes' unfolding method to datasets simulated with three hadronic interaction models and the composition model \cite{T. K. Gaisser} proposed by Gaisser, while using the response matrix and acceptance function constructed from a dataset simulated with the EPOS-LHC interaction model and the Shibata composition model \cite{M. Shibata}.

The unfolded spectra of the EPOS-LHC/Gaisser dataset show deviations from the input flux within $\pm$10\%, except for a few data points at the highest energies, where the uncertainty is dominated by Monte Carlo statistics.  
This implies that the uncertainties arising from the unfolding technique and composition model assumptions contribute to a flux uncertainty of up to $\pm$10\%.

It was also found that the unfolded all-particle spectrum remains within $\pm$10\% of the input flux,  
even when the hadronic interaction models used to generate the analyzed datasets and to construct the response matrix and acceptance differ.

On the other hand, the unfolded spectra of individual mass groups exhibit a clear dependence on the choice of hadronic interaction model.  
In the case of the proton and helium spectra, this dependence amounts to approximately $\pm$25\% below $10^{6.5}$\,GeV.  
The reconstruction of the carbon flux also shows an uncertainty at the $\pm$25\% level below $10^{6}$\,GeV, increasing up to $\pm$50\% and diverging beyond $10^{6.5}$\,GeV.  
For the iron spectrum, the model dependence is significant even at lower energies, reaching +55\% and -30\% in the energy range of $10^{5.1}$\,GeV to $10^{6.7}$\,GeV.

The Monte Carlo statistics used in this study correspond to an actual observation time of 2.65\,days of the Tibet experiment.  
Since the statistical errors are smaller than the systematic uncertainties below approximately $10^{6}$\,GeV,  
a small dataset from the Tibet experiment is sufficient to determine the energy spectra of individual mass groups.  
Notably, the all-particle spectrum shows minimal dependence on the interaction model.  
Although the reconstructed spectra of individual mass groups still show some level of model dependence,  
an energy overlap with modern direct measurements around $10^{5}$\,GeV offers a valuable opportunity to test hadronic interaction models.

In the energy range above $10^{6}$ GeV, our results show increasing model dependence, reaching deviations of more than $\pm$50 \%. This is partly attributable to the limited statistics of the Monte Carlo datasets.
%It is also explained by the fact that the separation of mass groups at the highest energies in this study is more difficult, as indicated in Fig.\ref{fig:sumrho-summu-contour}, and is possibly more model sensitive.
It is also explained by the fact that the separation of mass groups at the highest energies in this study becomes more difficult and is possibly more model sensitive. This is because the Tibet experiment being carried out at an altitude optimized to observe shower maxima around the knee energy, where the determination of the primary energy is nearly independent of the mass of the primary nucleus, thereby making the differences in secondary components such as $\Sigma \rho$ and $\Sigma (N_{\mu})_{31,\mathrm{m}}$ less discernible, as clearly demonstrated in Fig.~\ref{fig:sumrho-summu-contour}.

Enhancing the precision in this region is crucial for understanding the transition of the primary mass composition around the knee, one of the most important open questions in cosmic-ray physics. To achieve this, more extensive Monte Carlo event generations and greater experimental exposure will be required. A dedicated study targeting the knee energy region is now in progress, and further improvements are expected in future analyses.

%In the energy range above $10^{6}$ GeV, our results exhibit a larger model dependence, exceeding $\pm$50\%.  
%This is, however, partly due to the limited statistics of the Monte Carlo datasets, and improved performance is expected in future studies.  
%Enhancing the precision in this energy region is important for understanding the transition of the primary mass composition around the knee,  
%which remains one of the most significant questions in cosmic-ray physics.
%Increasing model-dependent deviations at the highest energy are partly due to the lack of MC statistics.
%In addition, as found in Fig.\ref{fig:sumrho-summu-contour}, the separation at the highest energy range becomes more difficult than at the
%lower energy and may be more model sensitive.  More careful study at this energy range requires more MC
%event generations and correspondingly more experimental exposure.  A dedicated study to cover the knee
%energy region is now ongoing.Increasing model-dependent deviations at the highest energy are partly due to the lack of MC statistics.
%In addition, as found in Fig.\ref{fig:sumrho-summu-contour}, the separation at the highest energy range becomes more difficult than at the
%lower energy and may be more model sensitive.  More careful study at this energy range requires more MC
%event generations and correspondingly more experimental exposure.  A dedicated study to cover the knee
%energy region is now ongoing.
%
\section{Acknowledgements} \label{Acknowledgements}

%The collaborative experiment of the Tibet Air Shower Arrays has been
%conducted under the auspices of the Ministry of Science and Technology
%of China and the Ministry of Foreign Affairs of Japan. 
This work was
supported in part by %a Grant-in-Aid for Scientific Research on
%Priority Areas from the Ministry of Education, Culture, Sports,
%Science and Technology, and 
 Grants-in-Aid for Science Research from
the Japan Society for the Promotion of Science in Japan (24H00220, 15H03655, 21H04464). 
%This work issupported by the National Natural Science Foundation of China under
%Grants No.~12227804, No.~12275282, No.~12103056 and No.~12073050, and
%by State Key Laboratory of Particle Astrophysics, Institute of High Energy
%Physics, CAS.
This work is also supported by the joint research
program of the Institute for Cosmic Ray Research (ICRR), the
University of Tokyo.

\end{document}